\documentclass[structabstract]{aa} 

\usepackage{graphicx}
\usepackage{txfonts}

\def\bp{$\beta$~Pic\ }
\def\bpic{$\beta$~Pic\ }

\def\L'{m$_{L'}$\ }

\def\pm{$^{+}_{-}$}

\begin{document}

\title{{Constrains on planets around \bp with Harps radial velocity data}\thanks{Based on observations collected at the
European Southern Observatory, Chile, ESO; runs XXXXX.}}
\subtitle{ }

\author{
A.-M.~Lagrange \inst{1}
 \and
K.~De Bondt \inst{1}
\and
N. Meunier \inst{1}
\and
M. Sterzik \inst{2}
\and
H. Beust  \inst{1}
\and
F. Galland  \inst{1}
}

\offprints{A.-M. Lagrange}

\institute{
  Institut de Plan\'etologie et d'Astrophysique de Grenoble,
  Universit\'e Joseph Fourier, CNRS, BP 53, 38041 Grenoble, France
  \email{anne-marie.lagrange@obs.ujf-grenoble.fr}
\and
  ESO, Karl Schwarzschild St, 2, D-85748 Garching 
}

\date{Received date / Accepted date: 24 01 2012}

\abstract
{ The $\beta$ Pictoris system with its debris disk and a massive giant planet orbiting at $\simeq$ 9 AU represents an
  ideal laboratory to study giant planet formation and evolution as well as planet-disk interactions. \bp b can also help testing brightness-mass relations at young ages. Other planets, yet undetected, may of course be present in the system.  }
{We aim at putting direct constrains on the mass of \bp b and at searching for additional jovian planets on orbits closer than typically 2 AU.}
{We use high precision Harps data collected over 8 years since 2003 to measure and analyse \bp radial velocities. }
{ We show that the true mass of \bp b is less than 
10, 12, 15.5, 20 and 25 MJup if orbiting respectively at 8, 9, 10, 11 and 12 AU.  This is the first direct constraint on the mass of an imaged planet. The upper mass found is well in the range predicted by brightness-mass relations provided by current "hot start" models. We also exclude the presence of giant planets more massive than 2.5 MJup with periods less than 100 days (hot Jupiters), more massive than 9 MJup for periods in the range 100-500 days. In the 500-1000 day range, the detection limit is in the brown dwarf domain. Beyond the intrinsic interest for \bp, these results show the possibilities of precise RV measurements of early type, rapidly rotating stars.
 }
{}


\maketitle

\section{Introduction}
Since the imaging of a circumstellar debris
disk in the eighties, the young ($12^{+8}_{-4}$~Myr; \cite{zuckerman01}) and close ($19.3 ^{+}_{-}0.2$~pc; \cite{crifo97}) A5V star $\beta$ Pictoris\, has been considered as a prototype of young planetary systems. Together with the other debris disks imaged since, this system allows to study the physical and chemical characteristics of sites of on-going or just finished  planetary formation. Recently, we were able to detect a companion orbiting the star at a distance ranging between 8 and 15 AU (\cite{lagrange09}). Its $L'=11.2$ approximate apparent magnitude translates into a temperature 
of $\sim 1500$~K and a mass of $\sim8$~M$_{Jup}$ according to Lyon's group models (\cite{baraffe03}).  Similar masses were later found from images at Ks (\cite{bonnefoy11}), and at 4 $\mu $ (\cite{quanz10}).
If orbiting on a slightly inclined orbit, the companion could explain most morphological (asymmetries) and dynamical peculiarities of the $\beta$ Pictoris\, dust system, as well as the "falling evaporating bodies" (FEB) phenomenon at the origin of the replenishment of the disk gas phase. In particular, it could explain the characteristics of the warp observed within the inner part of the disk (\cite{lagrange10} and ref. therein). Additional data now show that the planet semi-major axis is probably in the range 8-12 AU (see \cite{chauvin11} and references there-in). 
 Interestingly in the present context of planet formation theories, \bpic b is located in the region where, according to \cite{kennedy08} models, and given the star age and mass, giant planets can have formed {\it in situ} by core-accretion, in contrast to the few other planets imaged so far (\cite{marois08}, \cite{marois10}, \cite{kalas08}, \cite{lafreniere10}, \cite{chauvin05a}, \cite{chauvin05b}), located further away from their parents.

The mass determination of \bp b as well as those of  all other planets imaged relies however on the model dependant brightness-mass relation provided by the so-called "hot-start" models. In these models, the giant planet is formed by the collapse of a gaseous cloud and all the energy released is transfered in heat. The young planet is therefore quite hot during its first 100 Myr, which may allow planet direct detection around young stars with current imagers. Recently, however, \cite{fortney08} have developped an alternative model, expected to provide brightness-mass relationships in the case of core-accretion. In their model ("cold-start" model), a significant amount of energy is lost during the accretion process of gas onto a solid core. As a result of this loss of energy, their model predicts planets much fainter than the "hot-start" model predictions at young ages, at a given mass: for instance, a 10 Jupiter mass planet aged 10 Myr is $\simeq$ 5-6 magnitudes fainter. Under such assumptions, and except during the very short accretion phasis, young planets would not be detectable with current instruments. We note that the discrepancy between both models decreases with time, but remains significant up to 100 Myr. Coming back to the \bp b case, the observed magnitude is compatible with a planet formed under the "hot-start" model, but not under the "cold-start" model, as presently developped. In particular, this "cold-start" model does not predict any planet with such luminosity at 10 Myr. However, it would be prematurate to conclude that \bp b did not form by core-accretion based on these brightness-mass considerations, as these models, 1) probably represent two extreme situations in between which  actual processes at work during giant planet formation take place, and, 2) have not been calibrated so far by means of observations. Calibrations require independant measurements of the mass of the planets, ie dynamical masses, derived from astrometric or RV measurements. \footnote{Such masses measurements are refered in the following to as "direct", in contrast with model-dependent, brightness-based estimations.} This is why detecting and characterizing planets, in particular young ones, is so badly needed. 

Another important issue is the presence of other planets around $\beta$ Pictoris. The example of HR8799 (\cite{marois10}) and the radial velocities (hereafter RV) study of solar-type stars (http://www.exoplanet.eu) show that multiple planetary systems are indeed very common. Also, gaps have been reported in the \bp disk in the mid-IR (\cite{okamoto04}, \cite{wahhaj03}, \cite{telesco05}) that were sometimes attributed to planets 
(\cite{freistetter07}). Our available images do not show evidence of planets more massive than $\sim$ 5 M$_{Jup}$ at separations of  10-20 AU (\cite{lagrange10}). As current imagers are not able to put constrains on the presence of planets closer than typically 5-6 AU from the parent stars, other methods are needed to get information on the presence of planets in the inner 5 AUs. An analysis of 120 spectra obtained with Coralie in 1998-1999 as well as 258 Harps spectra obtained between November 2003 and July 2004 allowed us earlier to exclude the presence of an inner giant planet with masses larger than 2 M$_{Jup}$ at a distance to the star of 0.05 AU, and 9 MJup at 1 AU (\cite{galland06}). In this paper, we analyse more than 1000 Harps spectra recorded between 2003 and 2011 (described in Section 2) to search for additional, closer planets (Section 3) and to further constrain \bp b mass (Section 4).

\section{Data}
We have now gathered 1098 Harps spectra from the ESO Archive or obtained during our observing runs. These data span a period of time from October 2003 to February 2011.  Exposure times of individual spectra are between typically 60  and 120 seconds, and typical SNs vary between 150 at 250 at 400 nm. 

All data have been reduced and analysed with our SAFIR software (\cite{galland05}), that we developped to measure RV of early type stars precisely enough to detect planets. Such stars are usually rapid rotators and show much fewer lines than solar type stars; they are then not adapted to masks-based technics. The method consists in correlating, in the Fourier space, each spectrum and a reference spectrum built by summing-up all the available spectra for this star to measure its RV. 

\begin{figure}
\centering
\includegraphics[angle=0,width=\hsize]{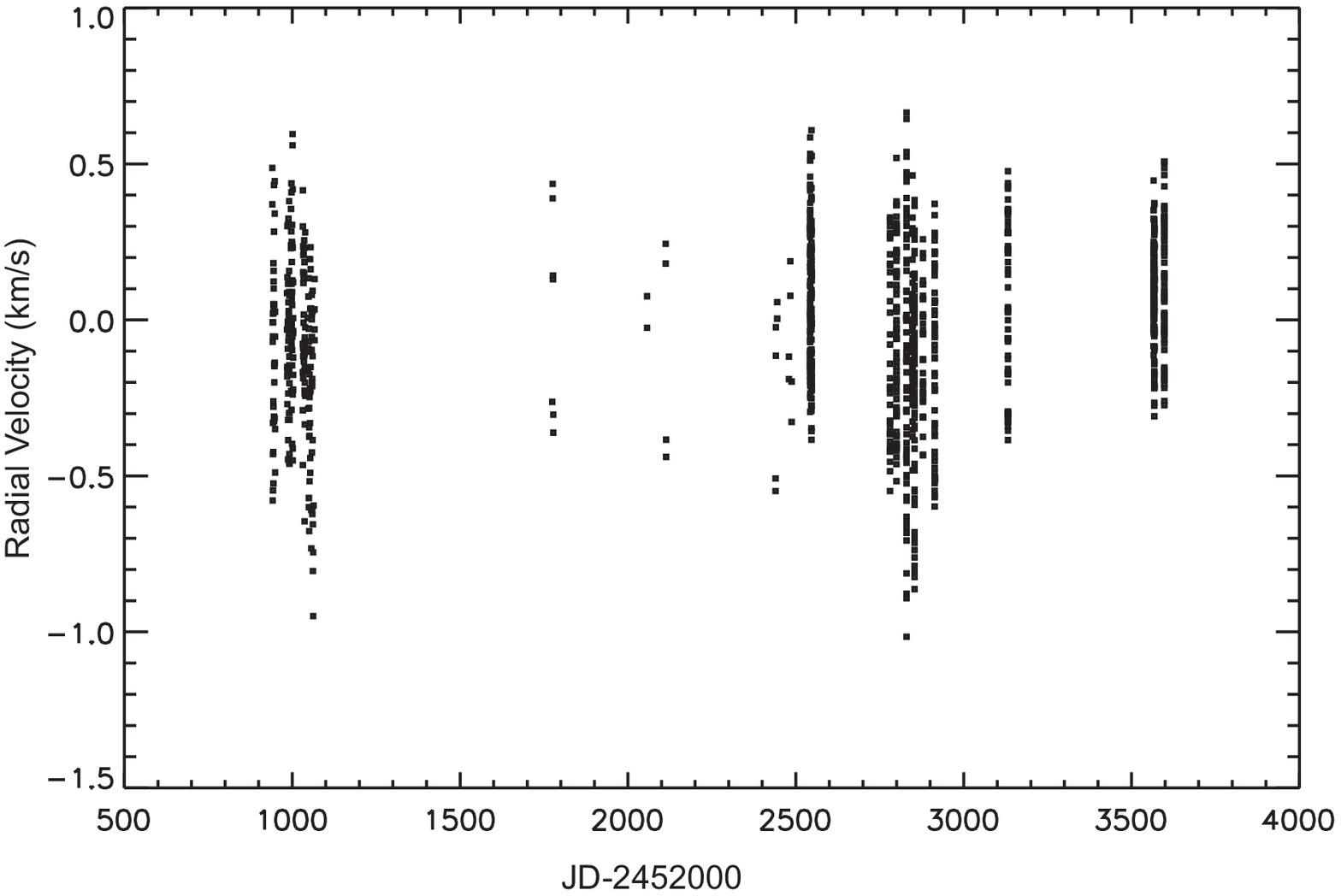}
\includegraphics[angle=0,width=\hsize]{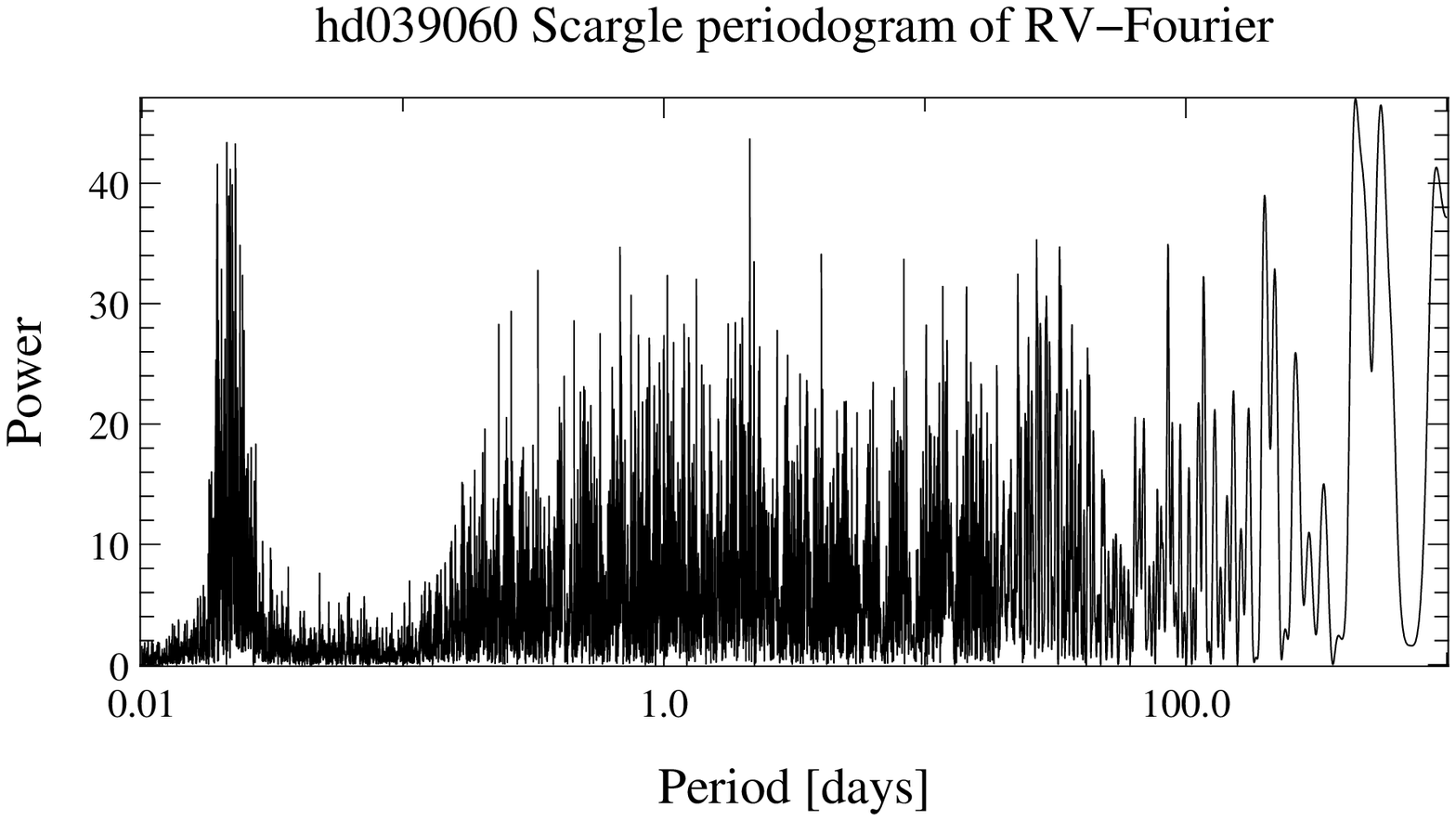}
  \caption{Top: 
RV data of \bp between 2003 and 2011. Bottom: associated periodogram.}
\label{RV_alldata.eps}
\end{figure}

\begin{figure}
\centering
\includegraphics[angle=0,width=\hsize]{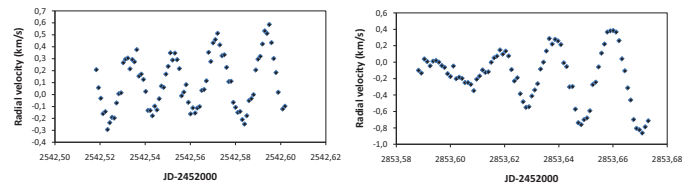}
  \caption{Examples of high frequency RV variations.}
 \label{RV_zoom.eps}
\end{figure}

\begin{figure}
\centering
\includegraphics[angle=0,width=\hsize]{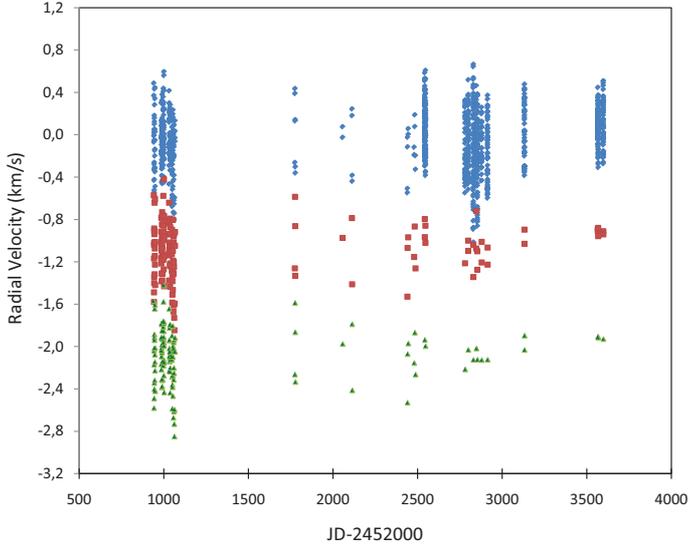}
  \caption{RV data of \bp between 2003 and 2011, averaged over 0.5 (red) and 1 (green) day. For comparison, the unaveraged data are also given in blue. Averaged data have been vertically shifted for clarity purpose.}
 \label{RV_averaged.eps}
\end{figure}

The RV thus obtained are shown in Figure~\ref{RV_alldata.eps}. The most obvious feature is the presence of high 
frequency (HF) variations, already described in Coralie and Harps spectra (\cite{galland06}) at 47.4\pm0.01 cycle/day (period
of 30.4 min) and 39.05\pm0.01 cycle/day (period of 36.9 min) and  part of a complex pattern of pulsations described in \cite{koen03}.  Before 2008, each pointing (refered to as a visit) consisted generally in two consecutive exposures, and lasted typically 5-10  minutes, including overheads. Starting from March 2008, we adopted an observational strategy aimed at attenuating these HF variations.  The duration of the visits was increased to $\simeq$ 2 hours, during which we recorded several 1-2 minutes consecutive exposures, to average out as much as possible the HF variations. Examples of the RV obtained during these  2hour-visits are provided in Figure~\ref{RV_zoom.eps}. The period March 2008 (JD 2454542) until February 2011 (JD 2455597, referred to as "set1"; 810 spectra over $\simeq$ 1000 days) defines then a much more homogeneous set of data than the whole set of data. The impact of this strategy on the RV curve is illustrated in Figure~\ref{RV_averaged.eps} where we show the data once averaged over 0.5 day and 1 day. The improvement  is also clear when we consider the rms of the data (see Table~ \ref{stats}), which is significantly decreased once the data have been averaged.
 
\begin{table}
\caption{ rms (m/s) of all RV measurements between 2003 and 2011 and rms of the uncertainties (m/s) associated to the RV measurements.}
\label{stats}
\centering
\renewcommand{\footnoterule}{}  
\begin{tabular}{l l l l l l l l l l}
\hline 
\multicolumn{4}{c}{}     \\
 Data set &  
no average &  
average 0.5d &  
average 1d \\
\hline
all data & 275.2 (39.1) & 257.7 (27.3) & 248.1 (27.3)\\
\hline
set1 
&  277.6 (39.7)
& 97.2 (5.2) &97.2 (5.2)
\\
\hline
\end{tabular}
\end{table}

\section{ Search for short period planets}
\subsection{ rms-based detection limits}
Our SAFIR software provides estimations of the detection limits. These limits are computed in the following way: for a given planet (Mass, period), we compute the induced RV at the same observing dates as the real observations and the rms of these simulated RV, for several (typically 1000) values of orbital phase. The distribution of the obtained rms is gaussian. The planet is detected if the rms of the observed RV (that we call hereafter RV-jitter) is smaller than the average of the rms distribution of the simulated RVs. The level of confidence (detection probability) is computed by comparing the standard deviation of the simulated distribution and the gap between the observed rms and the average value of the simulated distribution.  
We showed in \cite{lagrange09} that when the signal is well sampled, the detection limit is close to the one that 
corresponds, for a given period, to a planet that would produce an RV amplitude equal to 3$\times$RV-jitter.
This computation of the detection limits is very fast; however, it does not take into account the temporal structure of the stellar noise. As the \bp variations are mainly HF variations (Figure~\ref{RV_alldata.eps}), 
this method tends to overestimate the detection limits for periods larger than one day (unless, when the data sampling allows it, temporal averaging can be done). We therefore developed in addition alternative methods that take into account the frequency-structure of the stellar noise. These approaches will be described in Sect. 3.2. 

\begin{figure}
\centering
\includegraphics[angle=0,width=\hsize]{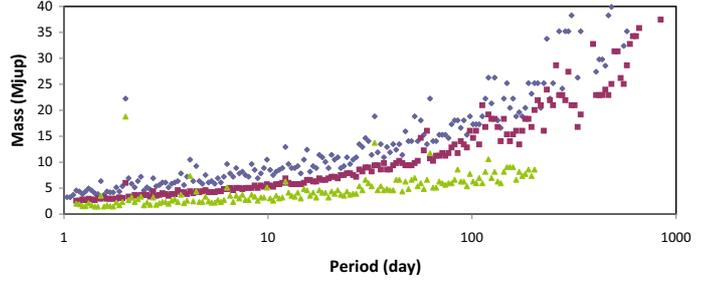}
 \caption{rms-based detection limits, for periods in the range 1-1000 days. Blue: all data, no average; red: all data, averaged over 1 day; green: set 1 data, averaged over 1 day.}
 \label{limdet_safir.eps}
\end{figure}

The obtained detection limits (99.7$\%$ probability) as computed on the whole set of data are provided in Figure~\ref{limdet_safir.eps}. Typical values obtained for P $\simeq$ 5, 10, 100, 500 and 1000 days are resp. 
7, 10, 16, 62 and 70 M$_{Jup}$ (all the computed masses given in this sub-section are rounded to the next integer or integer/2 mass). 
Note that as the system is seen edge-on, within at most a few degrees, if we assume that the planets orbit within the debris disk, then the detection limits derived are not affected by the sin(i) indetermination (i=90 degrees).

When averaging the RV over one day, we find detection limits of 4.5, 5.5, 13.5, 29 and 53 M$_{Jup}$ (see also Figure~\ref{limdet_safir.eps}). Hence, averaging over one day significantly improves the detection limits. This is consistent with the measured RV rms beeing dominated by the HF variations (see Table~\ref{stats}).

Finally, if we now consider set 1 data (no average), we get detection limits of 7, 8 and 16 M$_{Jup}$ for periods of 5, 10, 100 days,  ie comparable to the limits found using the whole set of data, when short periods are considered (limits for periods longer than typ. 200 days are not meaningful given the comparatively shorter duration of set 1). This similarity is due to the fact that the rms associated are comparable (see Table~ \ref{stats}). Finally, when averaging the data over one day, we find lower detection limits: 3, 4 and 6.5 M$_{Jup}$ for the same periods. These values are comparatively better than those obtained after averaging the whole set of data, eventhough the time span and the number of data are smaller. This illustrates the usefulness of our adopted observing strategy, which naturally averages the stellar jitter.

\subsection{ Periodogram-based detection limits}
Here we provide a brief description of alternative approaches we developped to take into account the structure of the stellar noise, and the associated results. We refer to a forthcoming paper {(Meunier et al, 2012, in prep)} for a detailed study of these approaches, and of their respective merits, depending on the available data.
\subsubsection{Highest peaks-based estimates} 
For a given planetary mass and a given period, we compute the RV induced by the planet at the same observing dates as for the real data. We add this planet-induced RV signal to the observed one. We then identify the five highest peaks in the planet periodogram. \footnote{In practice, the simulated periodogram is affected by the injection of the planet not only at the planet period but also close to the planet period: several peaks, induced by the planet, are present close to the planet period. Therefore, we have to exclude these peaks; to do so, we compute the periodogram corresponding to a planet with period=P$_{pla}$ for the same observing dates, and compute the associated median $m$ and standard deviation $s$ Meunier et al, 2012, in prep.). 
We then identify the peaks that are above a threshold of $m$+4$s$ and located within \pm 10\% P$_{pla}$ from P$_{pla}$ and we note Pf the period of the peak the furthest from P$_{pla}$. The window size is then defined by min(10\% P$_{pla}$; Pf).
}
The mean power of the five highest peaks outside the window defined in the footnote is then compared to the power of the planet peak. For each planet mass and period, we simulate 1000 realisations with different orbital phase between 0 and 2$\pi$. The planet is declared detectable if in 99.9 percent of the simulations, the planet peak is higher than the mean of the five highest peaks. Given the steps adopted, the calculated detection limit has an accuracy of 0.5 M$_{Jup}$ (set by the step taken in the computations)\footnote{Note: similar results were obtained using a criterion based on the fit of the RV data for periods corresponding to these five highest peaks. As the results are identical, we do not further describe this latter method in the present paper but refer to {Meunier et al, 2012, in prep.} for such a description. }.

\begin{figure}
\centering
\includegraphics[angle=0,width=\hsize]{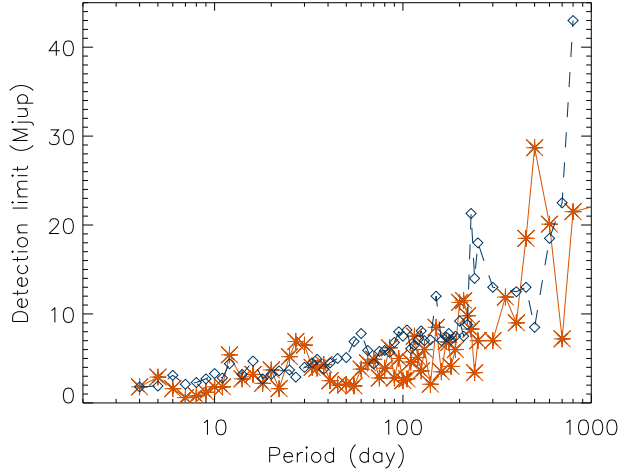}
 \caption{Detection limits found with the peak-based approach (5 peaks, blue dashed/diamonds), and the correlation approach (red full line/stars).}
\label{fig_limdet1b.ps}
\end{figure}

The detection limits obtained are shown in  Figure~\ref{fig_limdet1b.ps}.
Typical values obtained for P $\simeq$ 5, 10, 100, and 500 days are resp.  2.4, 3.8, 8.0, and 9.0 M$_{Jup}$ (99.9$\%$ probability), hence in general significantly better than the rms-jitter-based ones when using un-averaged data; the difference is not so important when compared to the set1, rms-jitter-based data averaged over 1 day. This is consistent with the reduced jitter obtained once averaging set1 data. 
For a 1000 day-period, the detection limit is larger than 50 M$_{Jup}$. 

\subsubsection{Correlation-based estimates}
\begin{figure}
\centering
\includegraphics[angle=0,width=0.8\hsize]{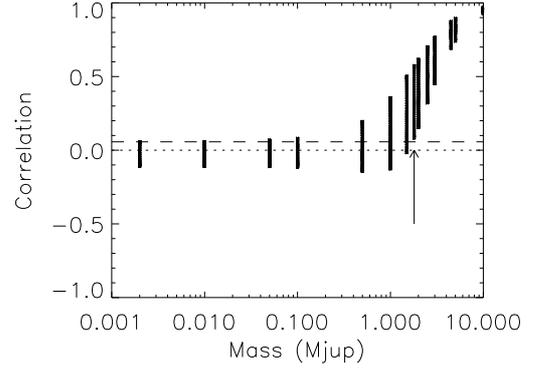}
  \caption{Correlations for 100 realisations at each planet mass versus the planet mass (in M$_{Jup}$), for a period of 10 days. The solid line shows the mean for each mass. The dashed line indicates the threshold (determined from the maximum correlation at very low planet mass) and the arrow shows the detection limit.}
 \label{Fig2_P10_correlation.eps}
\end{figure}

In a second approach, we compare the periodogram of the planet alone to the periodogram corresponding to the total radial velocities (ie the observed + planet induced RV). We compute the correlation between those two periodograms again for 1000 realisations corresponding to the same planet mass, and the same planet period, and 1000 different orbital phase. 
In Figure~\ref{Fig2_P10_correlation.eps}, 
the obtained correlation values are shown as a function of the planet mass, for a given period (here 10 days). We see that for low planetary masses, the upper limits of the correlations for a given planet mass remain constant (ie do not depend on the planet mass). So do the maximum and minimum values at each mass. At these low masses, the planet is certainly not detected and so the correlation values there correspond to a 'no detection'-case.  We define the detection limit as the lowest planet mass for which the smallest correlation value is higher than the threshold determined by the maximum correlation at the very low mass of 0.002 M$_{Jup}$. Hence the detection limit is the lowest mass for which 99.9$\%$ of the realizations have a correlation above that threshold.

The detection limits obtained are shown in  Figure~\ref{fig_limdet1b.ps}.
  Typical values obtained  are resp. 2.9, 1.8, 2.5, and 28.7 M$_{Jup}$ for P $\simeq$ 5, 10, 100, and 500 days (99.7$\%$ probability). Again they are better than the rms-based detection limits (no averaging), and more comparable (and actually slightly better)  to those obtained by the highest peaks-based estimates. 


\subsection{ Planets with periods smaller than 1000 days: conclusions}
From the above study, we  conclude that there is no planets with masses larger than  2.5 M$_{Jup}$ and  periods shorter than 100 days around \bpic . This excludes then the presence of massive hot Jupiters around \bp, whereas the star has certainly been surrounded by a massive protoplanetary disk and at least one outer planet has formed out of this disk. For planets with periods in the range 100-500 days, the detection limit is in the range 2.5-9 M$_{Jup}$. Finally, for planet-periods the 500-1000 day range, the present detection limit is in the brown dwarf domain. These are the best detection limits found around this star but also around any early type, rapidly rotating star, and these results clearly show the  tremendous, yet sofar largely underestimated, possibilities of precise RV measurements of such stars. As far as \bp is concerned, more RV data will help further constraining the detection limit at periods in the range 100-1000 days, and of course, also constraining the presence of planets at even larger periods.  The present study shows that given the temporal structure of the stellar noise, 2 hour-visits per year can in the future allow to improve significantly the present detection limits for periods $\geq$ 200 days. Evenmore, planets with periods as large as 1000 days or more could be tested. Coupled with other observing technics such as interferometry (\cite{absil11}), aperture masking (\cite{lacour11}) and AO imaging (\cite{lagrange10}), this would allow to get for the first time a complete coverage of the mass-separation plane for giant planets and for separation as small as a few tenths of AU to several AU.

\subsection{Direct constrains on \bp b mass}
The present data cover only 8 years, and do not therefore allow to estimate detection limits for periods longer than typically 1000 days, ie further than 2.5 AU. However, these data can be very usefull to constrain \bp b mass as we have additional  information on the planet position between 2003 and 2011. According to present knowledge, the planet semi major axis is probably between 8 and 12 AU, with the most probable orbit at 9 AU. Its eccentricity is small {($\leq$ 0.17)} and its longitude of periastron (computed for instance from the line of sight) is not well constrained. 
Very importantly, in 2003, \bp b was close to a quadrature, moving away from us to pass behind the star, and in 2011, close to the other quadrature, coming towards us (\cite{lagrange10}; \cite{chauvin11}). This is a very favorable situation as we can expect \bp RV variations induced by \bp b to be close to the maximum amplitude at both dates.  Knowing the companion positions in 2003 and 2011, using the fact that its orbital plane is seen edge-on, the RV data provide us with direct constrains on its mass, conversely to imaging which give model dependent (through brightness-mass relationships) estimates. 

To quantify an upper limit on \bp b mass, we proceed as follows:
\begin{itemize}
\item[-] First, we measured the slope of the observed RV; the value found is 3.1e-5 km/s/day, {with a fitting error of 1.4 e-6 km/s/day. To estimate the actual error on the slope, we } performed 1000 bootstraps of the data and measured the slopes obtained for each bootstrap. The average of the 1000 slopes is found to be 5.7 e-8 km/s/day (i.e. close to 0, as expected), and the dispersion is $\sigma$=1.02e-5 km/s/day, {a value unsurprisingly much larger than the error associated to the fitting process}. We note that, as expected, this dispersion is identical to that obtained on 1000 realisations of random noise with an rms of 275m/s with the same calendar. This shows that the measured slope is detected at a level of 3 sigma. 
\item[-] Then, we computed the detection limit for a planet with orbital characteristics similar to those of \bpic b. 
We assume that the planet was at quadrature beginning of January 2003, the currently most probable date (\cite{chauvin11}). We consider then a planet with a given semi-major axis (we assume a circular orbit), and a mass between 0.1 and 25 M$_{Jup}$ (steps of 0.1 M$_{Jup}$), and compute the RV induced by this planet on a 1.75 solar mass star, at the dates of the actual data.  We add random noise with an rms of 275 m/s (corresponding to the jitter measured on \bp RV data), and we perform a linear fit of these simulated RV, taking into account the error bars measured at each date on \bp data, to estimate the slope of the variations together with associated error bars. An example of the RV variations is showed in  Figure~\ref{fig_30MJup} in the case of a 30 M$_{Jup}$ planet located on a 9 AU (P=7400 d period): eventhough the sinusoidal shape of the variations is not obvious due to the high level of jitter, a significant positive drift is observed, which is not present in the \bp RV (Figure~\ref{RV_alldata.eps}). For each planet mass, we repeat this operation for 1000 different noise realisations in order to fully take into account the impact of the jitter noise. The values of the slopes thus obtained are showed in Figure~\ref{slopes} for a 9 AU orbit (in this figure, for the sake of clarity we have taken steps of 0.5 M$_{Jup}$ instead of 0.1M$_{Jup}$). We consider that a planet is detected if in 99$\%$ of the realisations, the slopes obtained are larger than 3$\times$$\sigma$= 3$\times$1.02e-5 km/s/day.\footnote{Doing this, we assume that the level of dispersion in the slopes resulting from the boostrap of the \bpic RV data is entirely due to stellar noise, which is a conservative estimation, as the observed slope indicates a drift at a 3 sigma level.} 
\item[-] We then explored the impact of the date of quadrature, assuming a range of -70;+130 days with respect to the most probable date, following \cite{chauvin11}.  The impact on the planet mass is found to be 0.5-2 M$_{Jup}$, for semi-major axis in the range 8-12 AU.
\end{itemize}
 Table~\ref{bpicbmass} give for semi-major axis of 8, 9, 10, 11, and 12 AU, the obtained upper limits.

\begin{table}
\caption{Constraints on \bp b mass. See text.}
\label{bpicbmass}
\centering
\renewcommand{\footnoterule}{}  
\begin{tabular}{l l l l l l l l l l}
\hline 
semi-major axis (AU) & 8&9&10&11&12\cr
upper mass (M$_{Jup}$)  &9.6 &12.0 &15.4  & 20.0 &24.8\cr
\hline
\end{tabular}
\end{table}

 To try to improve these upper limits, we performed a similar analysis using data averaged over 10 and 30 days. The upper limits found are not improved which is due to the fact that because of  not optimal sampling of the data taken prior to 2008, the jitter is still high (100 m/s or more).

\begin{figure*}
\centering
\includegraphics[angle=-90,width=0.45\hsize]{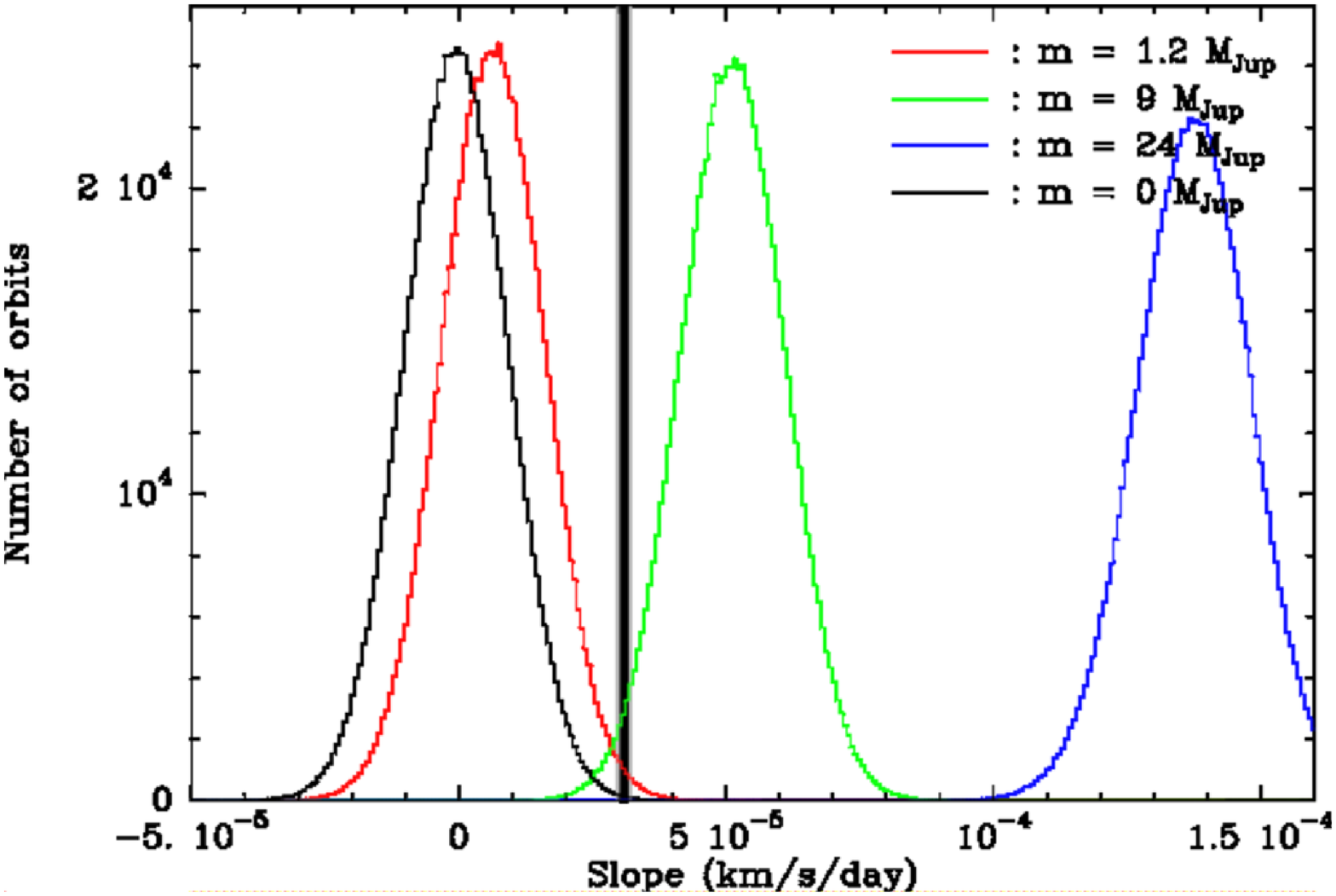}
\includegraphics[angle=-90,width=0.45\hsize]{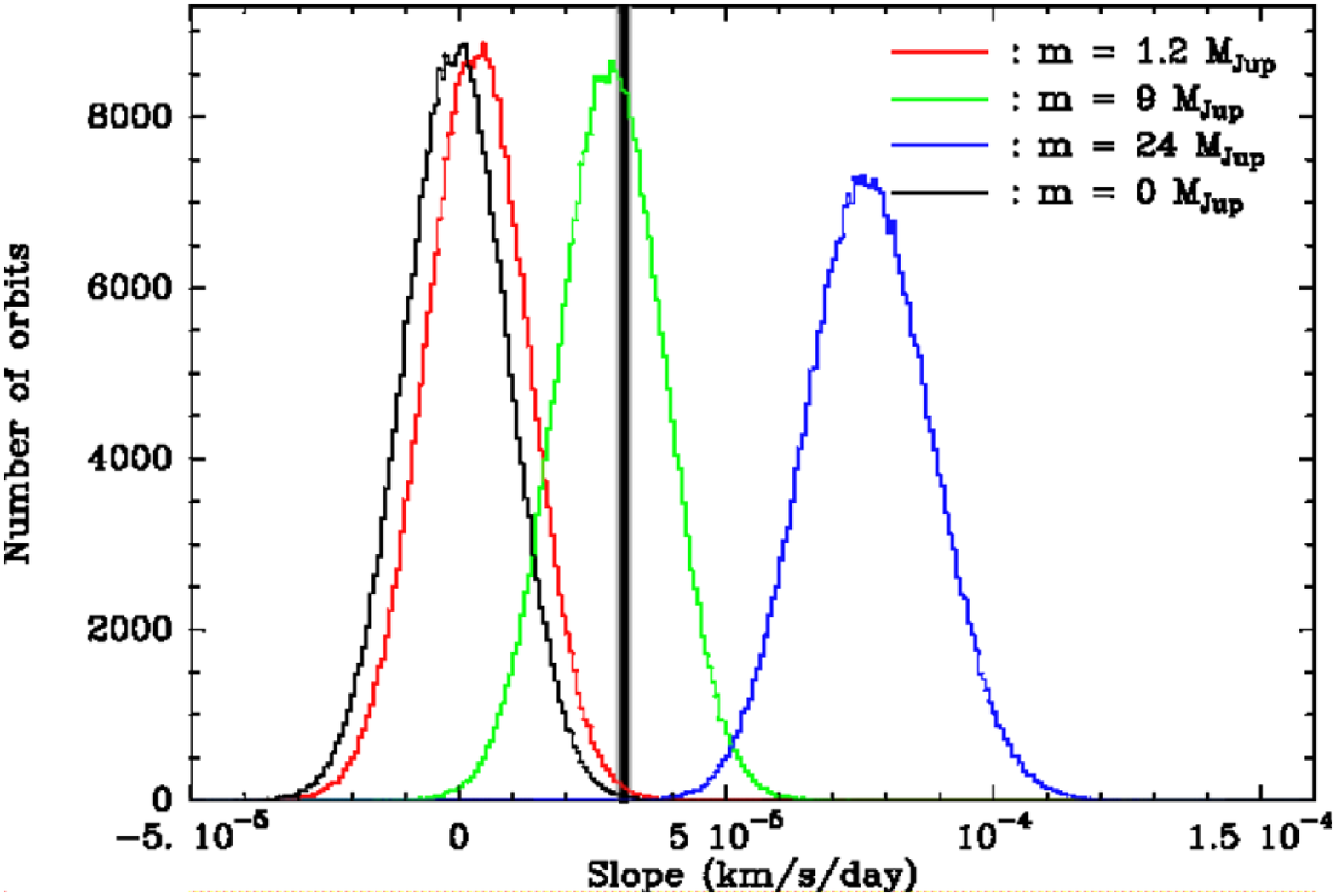}
  \caption{Plots of the statistical distribution of the simulated radial velocity slope using the model developped in Chauvin et al (2011), for three different masses : 1.2 MJup (red line), 9 MJup (green line) and 24 MJup (blue line) and semi-major axis in the range 8.5-8.5 AU (left) and 11.5-12.5 AU (right). The vertical grey/black bar shows the measured slope value, which is also the 3$\sigma$ upper limit of the slope.}
 \label{simus}
\end{figure*}

\begin{figure}
\centering
\includegraphics[angle=0,width=\hsize]{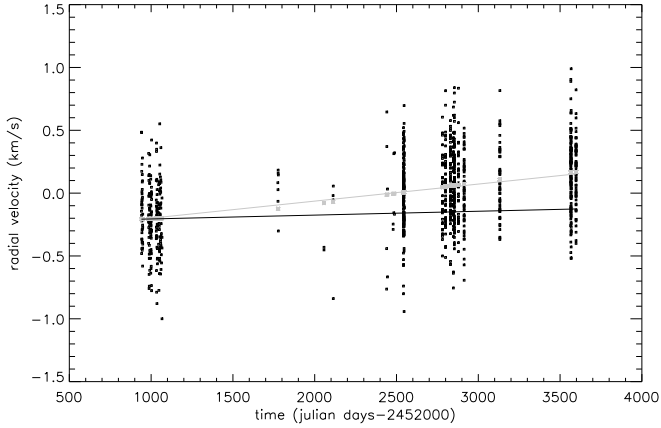}
 \caption{RV induced by a 30 M$_{Jup}$ planet with a 7400 day-period (9 AU sep.) and located at the position of \bp b in 2003, when using the actual  data sampling: grey crosses: without jitter noise, black squares: with a noise of 275 m/s. The grey line indicates the slope derived from the linear fitting of RV variations, and the black line indicates the slope of \bp real data deduced from linear fitting.   }
\label{fig_30MJup}
\end{figure}

\begin{figure}
\centering
\includegraphics[angle=0,width=\hsize]{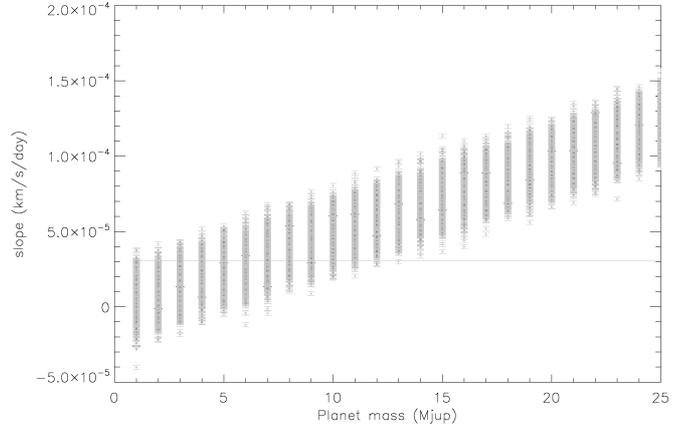}
 \caption{Slopes of the RV induced by a planet with different masses with a 7400 day-period when using the actual data sampling (steps of 0.5M$_{Jup}$); the horizontal lines indicate the 3sigma upper level derived from \bpic  data. }
\label{slopes}
\end{figure}

 In order to test the robustness of these results, as well as the impact of the planet eccentricity, we considered  the statistical orbit distribution obtained from our MCMC analysis in Chauvin et al. (2011). For each model in this distribution, we computed the expected radial velocities at the observing dates, assuming a given planetary mass and eccentricities up to 0.17 {following \cite{chauvin11}}. We then computed the statistical distribution of the slopes after fitting the RV curve with a linear fit. The slope depends of course on the assumed mass of Beta Pic b. This allows us a compatible mass range, assuming that the measured slope must fall within the computed distribution; the slopes are then compared to the 3$\sigma$ limit. The results are found to be in very good agreement with those deduced from the previous analysis. This is illustrated in Fig~\ref{simus} where the slope  distribution is plotted for 3 masses (1.2, 9 and 24 MJup) and in the case of semi-major axis in the ranges 8.5-9.5 AU (left) and 11.5-12.5 AU (right). 
 In the left figure, corresponding then to a $\simeq$ 9 AU, the distribution of slopes for a 24 MJup mass is well shifted compared to the 3$\sigma$ limit, where as the one corresponding to a 9 MJup is closer. For a value of $\simeq$ 12 AU, the  slope distribution with a 24 MJup mass is closer to the 3$\sigma$ limit.

{ Finally, we address the question whether these data also give constraints on the minimum mass of the planet. Assuming that the minimum slope is indeed equal to 3e-5 km/s/day, we can, as an exploratory exercice, derive a tentative lower limit to the mass of the planet if responsible for the slopes. Proceeding with a similar reasoning as above, and exploring again different dates of quadratures, we find lower masses in the range 1-2 MJup for semi-major axis between 8 and 12 AU. These conclusions are also compatible with the second approach described above, as distributions for e.g. 1.2 MJup are below the measured slope (see Fig.~\ref{simus}). The present values should be considered however only as an illustration of the potential of the present approach rather than definite results because the slope is so small and close to the limit; more data, with optimized temporal sampling are necessary to derive more reliable estimates of the \bp minimum mass. }

\subsection{Conclusion on \bp b mass }
The present data have then allowed to constrain directly the \bp b  mass, to be less than 
10, 12, 15.5, 20.0 and 25 MJup if orbiting respectively at 8, 9, 10, 11 and 12 AU. For the most probable orbit, 9 AU, the upper limit is in the planetary mass regime, which means that definitively, the source detected is not a more massive object (brown dwarf).  
This is the first time an imaged extra-solar planet mass is directly constrained, and thus does not depend only on model-dependant brightness-mass relationships. \bpic b upper mass falls quite in the range predicted by the brightness-mass relations provided by the "hot start" models. On the contrary, current cold-start models  fail to explain a giant planet with the observed brightnesses at Ks (\cite{bonnefoy11}), L' (\cite{lagrange10}) and M (\cite{quanz10}) bands. In the future, especially during the next coming years, new data with optimized sampling should allow to improve the mass range of \bp b. The present combination of RV and high contrast data on exoplanets is a pioniering one; it will be more generalized in the future, when more young planets will be imaged at short separations from their stars, and will give thus a precious opportunity to test brightness-mass predictions.


\begin{acknowledgements}
We acknowledge financial support from the French Programme National de Plan\'etologie (PNP, INSU). We also acknowledge support from the French National Research Agency (ANR) through project grant ANR10-BLANC0504-01. We also thank Julien Milli for fruitfull discussions {and the referee for his/her comments}.
\end{acknowledgements}

\end{document}